\documentclass[onecolumn,superscriptaddress,nofootinbib,notitlepage,longbibliography,aps]{revtex4-1}
\pdfoutput=1

\usepackage{graphicx}
\usepackage{amsmath}
\usepackage{amssymb}
\usepackage{amsfonts}
\usepackage[dvipsnames]{xcolor}
\usepackage[linktoc=none]{hyperref}
\usepackage[utf8]{inputenc}

\hypersetup{colorlinks=true,linkcolor=blue,citecolor=teal,filecolor=magenta, urlcolor=cyan}

\newcommand{\INFN}{INFN - Sezione di Napoli, Complesso Univ. Monte S. Angelo, I-80126 Napoli, Italy}
\newcommand{\UNINA}{Dipartimento di Fisica ``Ettore Pancini'', Università degli studi di Napoli ``Federico II'', Complesso Univ. Monte S. Angelo, I-80126 Napoli, Italy}
\newcommand{\SSM}{Scuola Superiore Meridionale, Università degli studi di Napoli ``Federico II'', Largo San Marcellino 10, 80138 Napoli, Italy}
\newcommand{\IFIC}{IFIC - Instituto de Física Corpuscular (CSIC-Universitat de València), c/ Catedrático José Beltrán 2, E-46980 Paterna (Valencia), Spain}

\begin{document}

\title{Constraints on heavy decaying dark matter with current gamma-ray measurements}

\author{Marco Chianese}
\email{chianese@na.infn.it}
\affiliation{\UNINA}
\affiliation{\INFN}
\author{Damiano F.G. Fiorillo}
\email{dfgfiorillo@na.infn.it}
\affiliation{\UNINA}
\affiliation{\INFN}
\author{Rasmi Hajjar}
\email{rasmienrique.hajjarmunoz@unina.it}
\affiliation{\SSM}
\affiliation{\color{black}\IFIC}
\author{Gennaro Miele}
\email{miele@na.infn.it}
\affiliation{\UNINA}
\affiliation{\INFN}
\affiliation{\SSM}
\author{Ninetta Saviano}
\email{nsaviano@na.infn.it}
\affiliation{\INFN}
\affiliation{\SSM}

\date{\today}
\begin{abstract}
Among the several strategies for indirect searches of dark matter, a very promising one is to look for the gamma-rays from decaying dark matter. Here we use the most up-to-date upper bounds on the gamma-ray flux from $10^5$ to $10^{11}$ GeV, obtained from CASA-MIA, KASCADE, KASCADE-Grande, Pierre Auger Observatory, Telescope Array {\color{black}and EAS-MSU}. We obtain global limits on dark matter lifetime in the range of masses $m_\mathrm{DM}=[10^7-10^{15}]~\mathrm{GeV}$. We provide the bounds for a set of decay channels chosen as representatives. The constraints derived here are new and cover a region of the parameter space not yet explored. We compare our results with the projected constraints from future neutrino telescopes, in order to quantify the improvement that will be obtained by the complementary high-energy neutrino searches.
\end{abstract}

\maketitle

\section{Introduction \label{sec:intro}}

Although Dark Matter (DM) is one of the cornerstones of the standard cosmological model, our effort to understand its nature has not come to fruition, leaving it as one of the most intriguing current unknowns of the composition of the Universe. Numerous attempts to directly observe DM have been performed, but only its gravitational effects have been observed so far. A complementary method to search for the existence of DM  consists in looking at the possible products of its decay or annihilation, that could leave significant imprints on the different astrophysical fluxes. This branch of research benefits from the existence of numerous telescopes providing  sensitivity over an enormous range of energies to exotic sources of Standard Model (SM) particles, in particular cosmic-rays, neutrinos, and photons.
 
Among the present and future telescopes we mention in the neutrino sector IceCube~\cite{Aartsen:2013jdh,Abbasi:2020jmh}, which introduced the neutrino messenger into the game, RNO-G~\cite{Aguilar:2020xnc}, GRAND~\cite{Alvarez-Muniz:2018bhp} and IceCube-Gen2 radio array~\cite{Aartsen:2019swn, Aartsen:2020fgd}, which aim to measure the very-high-energy neutrino fluxes. For cosmic-ray flux measurements we cite PAMELA~\cite{Picozza:2006nm}, AMS-02~\cite{Aguilar:2016kjl, Aguilar:2019owu} and the Pierre Auger Observatory (PAO)~\cite{Abraham_2010}. Finally, concerning the gamma-ray flux we find the measurements of Fermi-LAT~\cite{Ackermann:2014usa, Pieri:2009je}, CASA-BLANCA~\cite{Cassidy:1997sb}. Particularly relevant for the present work are the latest ultra-high-energy (UHE) results by CASA-MIA~\cite{CASA-MIA:1997tns}, KASCADE~\cite{KASCADEGrande:2017vwf}, KASCADE-Grande~\cite{KASCADEGrande:2017vwf}, PAO~\cite{PierreAuger:2015fol, PierreAuger:2016kuz}{\color{black},} TA~\cite{TelescopeArray:2018rbt} {\color{black} and EAS-MSU~\cite{Fomin:2017ypo}}. Very recently, the Tibet-AS$\gamma$ collaboration also published its gamma-ray measurements~\cite{TibetASgamma:2021tpz}.

In this work we focus on heavy DM, in the mass range from $10^7$~GeV to $10^{15}$~GeV, studying the decay of DM particles into a pair of SM particles with 100\% branching ratio within the minimal and model-independent decaying DM scenario with only two parameters: the DM mass $m_\mathrm{DM}$ and the lifetime $\tau_\mathrm{DM}$. We focus on the gamma-ray products of the DM decay chain, more concretely on the prompt contribution of gamma-rays, using the recent code \texttt{HDMSpectra}~\cite{Bauer:2020jay} to generate the $\gamma$-ray spectra. In this framework a dedicated analysis by Ishiwata et. al.~\cite{Ishiwata:2019aet} has obtained the limits on the lifetime of DM decaying into $b \bar{b}$ channel. However, the production of gamma-rays can be substantially different for {\color{black}different} decay channels. This is especially true for neutrinophilic decays, in which gamma-ray production only happens via the electroweak cascade following the decay. In this work, we use current high-energy gamma-ray measurements~\cite{CASA-MIA:1997tns,KASCADEGrande:2017vwf,PierreAuger:2015fol,PierreAuger:2016kuz,TelescopeArray:2018rbt} to perform a systematic study of models with different decay channels (see Refs.~\cite{Murase:2012xs,Esmaili:2012us,Feldstein:2013kka,Esmaili:2013gha,Rott:2014kfa,Boucenna:2015tra,Esmaili:2015xpa,Murase:2015gea,Cohen:2016uyg,Kalashev:2016cre,IceCube:2018tkk,Kachelriess:2018rty,Sui:2018bbh,Bhattacharya:2019ucd,Chianese:2019kyl,Dekker:2019gpe,Arguelles:2019boy,Kalashev:2020hqc,Esmaili:2021yaw,Maity:2021umk} for earlier gamma-ray and neutrino studies). In particular, we investigate benchmark examples for leptonic, neutrino, hadronic, and gauge boson channels. This extension is a necessary step towards the construction of uniform bounds on the parameter space of decaying DM, which will receive definite improvements with the information from future neutrino radio telescopes~\cite{Chianese:2021htv,Guepin:2021ljb}.

The paper is organized as follows. In section~\ref{sec:GammaFlux}, we describe the gamma-ray production from DM in the galactic halo and the absorption to which the gamma-rays are subject on their paths towards the Earth. We also show the experimental limits and the integral $\gamma$ fluxes that are obtained from the decay of galactic DM. In section~\ref{sec:ConsCons} we explain the statistical method we use in order to place the DM lifetime constraints, we show our results for the diverse decay channels, and we comment on the differences between the current gamma-ray limits and the projected neutrino ones. Finally, in section~\ref{sec:conclusions} we draw the main conclusions of this work.

\section{Gamma-ray flux from heavy dark matter}
\label{sec:GammaFlux}

One of the possible sources of UHE gamma-rays is the decay of heavy DM particles. The total gamma-ray flux expected from DM consists of different contributions stemming from prompt and secondary emission from galactic and extragalactic space. 
In this work we take into account only the prompt galactic emission which is the dominant component at the highest gamma-ray energies attained from DM decays, i.e. $E_\gamma = \mathcal{O}(m_\mathrm{DM}/2)$ with $m_\mathrm{DM}$ being the DM mass. We neglect both the extragalactic contribution, which is strongly suppressed at ultra-high energies by absorption due to $\gamma\gamma$ scattering, and the secondary galactic contribution originating from Inverse Compton scattering of electrons and positrons off low-energy photons (see, e.g., Ref.~\cite{Ishiwata:2019aet}). This component is in general subdominant to the galactic prompt one, and we have explicitly verified that it does not significantly influence our results. As will be clear later, this straightforward approach is appropriate to place global gamma-ray constraints on DM by means of gamma-ray observations which span several orders of magnitude in energy, from $10^5$ to $10^{11}$ GeV.

The expected prompt gamma-ray flux coming from galactic DM particles is~\cite{Esmaili:2015xpa}:
\begin{equation}\label{eq:prompt}
    \frac{\mathrm{d}\Phi_\gamma}{\mathrm{d}E_\gamma\mathrm{d}\Omega}(E_\gamma,b,l)=\frac{1}{4\pi\, m_\mathrm{DM}\tau_\mathrm{DM}}\frac{\mathrm{d}N_\gamma}{\mathrm{d}E_\gamma}(E_\gamma)\int_0^\infty \rho_\mathrm{DM}[r(s,b,l)]e^{-\tau_{\gamma\gamma}(E_\gamma,s,b,l)}\,\mathrm{d}s~,
\end{equation}
where  $\tau_\mathrm{DM}$ if the lifetime of the DM particle, respectively. The quantity $\mathrm{d}N_\gamma/\mathrm{d}E_\gamma$ is the energy spectrum of photons generated from the decay of DM particles, obtained from the \texttt{HDMSpectra} code~\cite{Bauer:2020jay}. The quantity $\rho_\mathrm{DM}(r)$ describes the DM galactic distribution, for which we adopt the commonly used Navarro-Frenk-White distribution
\begin{equation}\label{eq:NFW}
    \rho_\mathrm{DM}(r) = \frac{\rho_s}{r/r_s(1+r/r_s)^2}~,
\end{equation}
where $r_s=24~\mathrm{kpc}$ and $\rho_s=0.19~\mathrm{GeV~cm}^{-3}$ for the Milky Way~\cite{Cirelli:2010xx}. It is a function of the galactocentric radial coordinate, which in turn is related to the line-of-sight $s$ {\color{black}coordinate}:
\begin{equation}\label{eq:rcoord}
    r(s,b,l) = \sqrt{s^2+R_\odot^2-2sR_\odot\cos b\cos l}~,
\end{equation}
where $R_\odot = 8.5~\mathrm{kpc}$ is the distance of the Sun to the galactic center and ($b$, $l$) are the galactic angular coordinates.

Finally, the term in the exponent ($\tau_{\gamma\gamma}$) corresponds to the total optical depth of photons to $\gamma\gamma$ collisions, expressed as a sum of two different contributions:
\begin{equation}
    \tau_{\gamma\gamma}(E_\gamma,s,b,l)=\tau^\mathrm{CMB}_{\gamma\gamma}(E_\gamma,s)+\tau^\mathrm{SL+IR}_{\gamma\gamma}(E_\gamma,s,b,l)\,.
\end{equation}The first term takes into account pair production on Cosmic Microwave Background (CMB) photons: due to the homogeneity and isotropy of the CMB photons, the optical depth is
\begin{equation}\label{eq:cmb}
    \tau^\mathrm{CMB}_{\gamma\gamma}(E_\gamma,s) = \frac{4T_\mathrm{CMB}\,s}{\pi^2E_\gamma^2}\int_{m_e}^{\infty}\varepsilon_c^3\sigma_{\gamma\gamma}(\varepsilon_c)\ln\left(1-e^{-\frac{\varepsilon_c^2}{E_\gamma T_\mathrm{CMB}}}\right) \mathrm{d}\varepsilon_c~.
\end{equation}
Here, $\varepsilon_c=\sqrt{\varepsilon E_\gamma(1-\cos\theta)/2}$ is the photon center of momentum, $T_\mathrm{CMB}=2.348\times 10^{-4}~\mathrm{eV}$ is the CMB photons temperature, and $\sigma_{\gamma\gamma}$ is the pair production cross section given by
\begin{equation}\label{eq:xsec_pp}
    \sigma_{\gamma\gamma}=\frac{\pi}{2}\frac{\alpha^2}{m_e^2}(1-\beta^2)\left[ (3-\beta^4)\ln\left( \frac{1+\beta}{1-\beta} \right) -2\beta(2-\beta^2) \right]~.
\end{equation}
Here, $\alpha$ is the fine-structure constant, $m_e$ is the electron mass, $\theta$ the angle between the momenta of photons, and $\beta=\sqrt{1-1/\mathfrak{s}}$ with $\mathfrak{s}=\frac{\varepsilon E_\gamma}{2m_e^2}(1-\cos\theta)=\varepsilon_c^2/m_e^2$.

The second contribution to the total optical depth takes into account pair production from the interaction of the gamma-rays with the starlight and infrared light (SL+IR). Assuming that the photon field is inhomogeneous but isotropic we can write 
\begin{equation}\label{eq:slir}
    \tau^\mathrm{SL+IR}_{\gamma\gamma}(E_\gamma,s,b,l)= \int_0^s\mathrm{d}s'\iint \sigma_{\gamma\gamma}(E_\gamma,\varepsilon)\,n_\mathrm{SL+IR}[\varepsilon,\mathbf{x}(s',b,l)]\,\frac{1-\cos\theta}{2}\,\sin\theta\,\mathrm{d}\theta\,\mathrm{d}\varepsilon
\end{equation}
where $n_\mathrm{SL+IR}$ is the photon bath density extracted from \texttt{GALPROP} code~\cite{galprop}. Differently from the homogeneous CMB photon field, which is homogeneous, the SL+IR density strongly depends on the position $\mathbf{x}(s,b,l)$ in our galaxy, rapidly decreasing as we move away from the galactic disk.

\begin{figure}[t!]
    \centering
    \includegraphics[width=\textwidth]{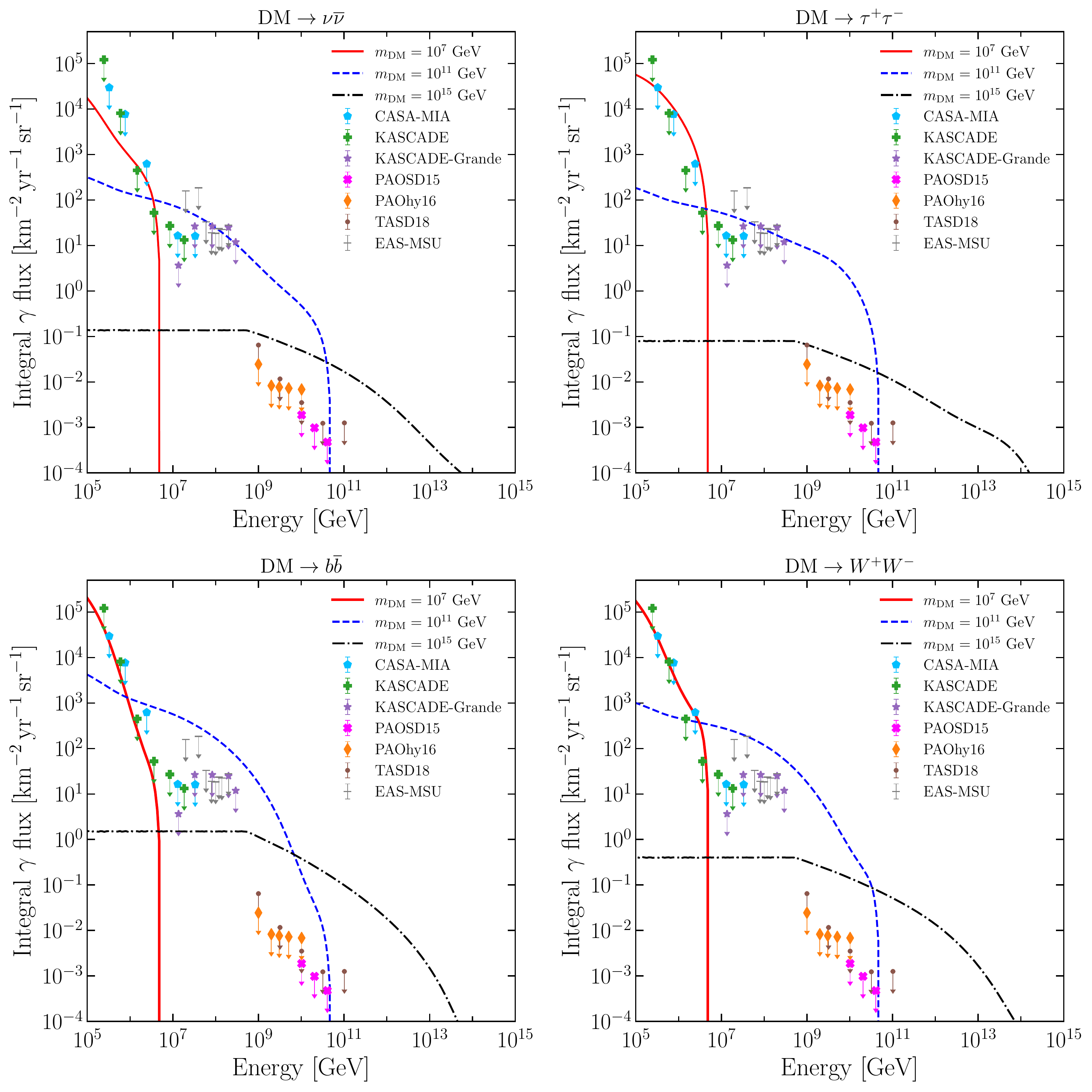}
    \caption{Integral $\gamma$-fluxes for the $\nu\bar{\nu}$, $\tau^+\tau^-$, $b\bar{b}$ and $W^+W^-$ decay channels assuming a lifetime of the DM particles of $\tau_\mathrm{DM}=10^{27}~\mathrm{s}$. These fluxes are obtained from equation~(\ref{eq:integral}) and we show them for three benchmark DM masses for each channel. We also include the upper limits on the gamma-ray flux placed by the collaborations. CASA-MIA, KASCADE{\color{black},} KASCADE-Grande {\color{black} and EAS-MSU} limits have a 90\% CL while the PAO and TA Surface Detector (TASD) ones have a 95\% CL.}\label{fig:intfluxes}
\end{figure}

In order to make a direct comparison with current limits on the diffuse UHE gamma-ray flux, from Eq.~\eqref{eq:prompt} we compute the angle-averaged integral $\gamma$-flux as
\begin{equation}\label{eq:integral}
      \Phi_\gamma(E_\gamma) = \frac{1}{4\pi} \int_{E_\gamma}^\infty \mathrm{d}E'_\gamma \int_{4\pi} \mathrm{d}\Omega \frac{\mathrm{d}\Phi_\gamma}{\mathrm{d}E'_\gamma\mathrm{d}\Omega}~.
\end{equation}
The average over all the directions $(b,l)$ of gamma-rays has also the effect of making our results robust against density profile choices. {\color{black}We have checked that the DM constraints obtained in this analysis weaken by less than 10\% in case of the Burkert density profile.}

In Fig.~\ref{fig:intfluxes} we show the integral $\gamma$-flux we obtain for a DM lifetime of $10^{27}~\mathrm{s}$ and different DM masses in case of four different DM decay channels. Also shown are the upper limits on the flux placed by different experimental collaborations: KASCADE~\cite{KASCADEGrande:2017vwf}, KASCADE-Grande~\cite{KASCADEGrande:2017vwf}, CASA-MIA~\cite{CASA-MIA:1997tns}, Pierre Auger Observatory (PAO)~\cite{PierreAuger:2015fol, PierreAuger:2016kuz}, and Telescope Array (TA)~\cite{TelescopeArray:2018rbt}. As can be seen in the plots, the experimental measurements are divided in two energy regions: CASA-MIA, KASCADE and KASCADE-Grande are sensitive to $E_\gamma \lesssim 10^9~\mathrm{GeV}$, whereas PAO and TA have measurements at higher energies. Moreover, we note that the lower DM masses have higher integral $\gamma$-fluxes concentrated in a small energy interval, while the higher DM masses have lower fluxes spread across a broader energy range. For $m_\mathrm{DM} = 10^{15}~\mathrm{GeV}$ (the highest DM mass we consider), the integral $\gamma$-fluxes reaches a plateau as the energy decreases. This is an artificial feature due to the truncation of the \texttt{HDMSpectra} spectra at $E_\gamma = 5 \times 10^{-7} m_\mathrm{DM} = 5 \times 10^{-9}~\mathrm{GeV}$, as well as to our neglecting of the secondary production. However, this has no impact on the global gamma-ray constraints, which are indeed mainly driven by PAO and TA data for high DM masses. This is the main reason why the results of our analysis are not changed by the introduction of the secondary galactic photons.

\section{Dark matter constraints} \label{sec:ConsCons}

In this section, we determine the constraints on the parameter space of heavy decaying DM from the current upper limits on the integral $\gamma$-flux. For each experimental data set shown in Fig.~\ref{fig:intfluxes}, we perform a $\chi^2$ analysis using the test statistic (TS) 
\begin{equation}\label{eq:xi2}
    \mathrm{TS}(m_\mathrm{DM},\,\tau_\mathrm{DM}) = \sum_{i=1}^{N}\left[\frac{\Phi_{\gamma,\,i}(m_\mathrm{DM},\,\tau_\mathrm{DM})-\Phi_{\gamma,\,i}^\mathrm{data}}{\sigma_i}\right]^2~,
\end{equation}
where $\Phi_{\gamma,\,i}$ is the expected DM integral $\gamma$-flux reported in Eq.~\eqref{eq:integral}, $\Phi_{\gamma,\,i}^\mathrm{data}$ are the experimental measurements with $\sigma_i$ being the corresponding uncertainty. The sum is over the total number of data points reported by each experiment. In particular, according to the definition of upper limits on the integral $\gamma$-flux, we simply have $\Phi_{\gamma,\,i}^\mathrm{data} = 0$. Moreover, we consistently compute the standard deviations $\sigma_i$ {\color{black} at 68\% CL} of each data point according to the confidence level of the upper limits provided by the experimental collaboration {\color{black} and reported in the caption of Fig.~\ref{fig:intfluxes}}. Hence, for a given DM mass from $10^7$ to $10^{15}$~GeV, we place a lower limit on the DM lifetime at 95\% CL by testing the hypothesis that gamma-rays were produced by decaying DM alone. {\color{black} In order to find the lifetime for which the observation of zero events is excluded with probability larger than 95\%, we impose $\mathrm{TS}=2.76$. To obtain this threshold value, we have determined the distribution of the test statistic under the assumption that the data came from DM decay. This distribution is not the usual chi-squared one, since the measurements are subject to the constraint that they must be larger than 0. Under such a constraint, the distribution becomes
\begin{equation}
    P(x)=\frac{e^{-x/2}}{\sqrt{2\pi x}}\frac{1+\Theta(\lambda-x)}{1+\text{Erf}\left(\frac{\lambda}{\sqrt{2}}\right)},
\end{equation}
where $x={\rm TS}$ and $\lambda=\sum_{i=1}^{N}\left[\Phi_{\gamma,\,i}(m_\mathrm{DM},\,\tau_\mathrm{DM})/\sigma_i\right]^2$, namely the TS corresponding to $\Phi_{\gamma,i}^{\mathrm{data}}=0$, and $\Theta$ denotes the Heaviside function. For $\lambda\to\infty$, corresponding to expected measurements much larger than 0, this distribution asymptotically reaches the usual chi-squared one, as it should. Since the measured TS coincides with $\lambda$, we have directly obtained the $95\%$ threshold value by taking $\lambda=2.76$.}

\begin{figure}[t!]
    \centering
    \includegraphics[width=\textwidth]{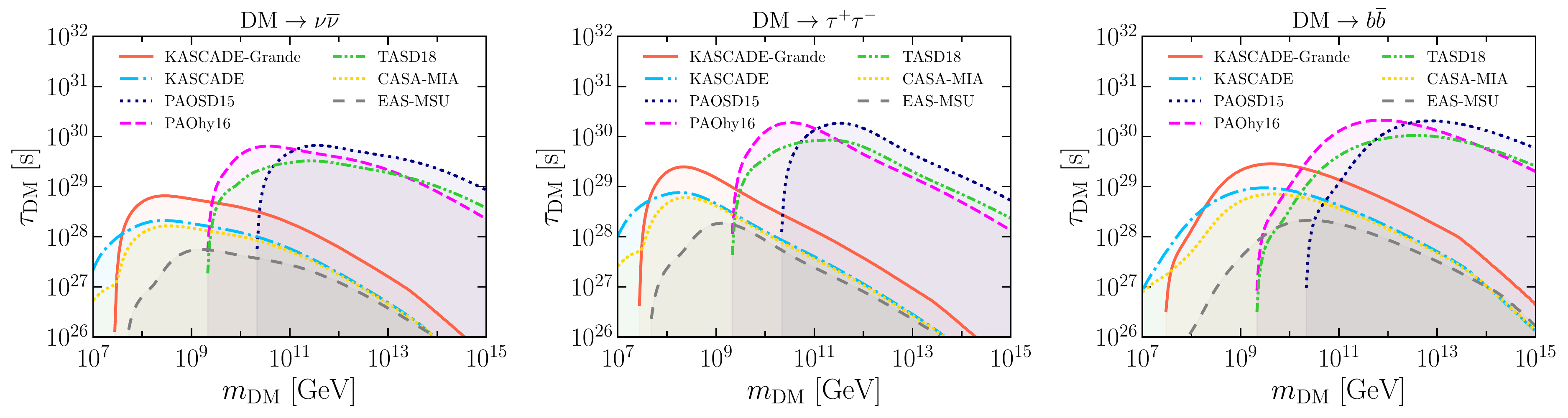}
    \caption{95\% CL lifetime limits of decaying DM particles for the $\nu\bar{\nu}$, $\tau^+\tau^-$ and $b\bar{b}$ channel using current measurements of the high-energy gamma-ray flux. Each panel corresponds to a decay channel, and for each channel we obtain the lifetime constraints using the upper limits provided by the experimental collaborations as shown in Fig.~\ref{fig:intfluxes}. The excluded regions are shown as shaded.}\label{fig:DM_life}.
\end{figure}

We present the main results of our analysis in Fig.~\ref{fig:DM_life}, where we show the exclusion limits at 95\% CL on the lifetime of heavy decaying DM for different decay channels. We find that the strongest limits are obtained from KASCADE (PAO) for DM masses smaller (larger) than $\sim 10^9$~GeV. All the constraints show a different low-mass cutoff below which the integral $\gamma$-fluxes contribute at energies smaller than the corresponding experimental sensitivities. Finally, a further comment on the \texttt{HDMSpectra} truncation of the photon spectra at low energies is in order. In particular, it marginally affects only the constraints set by CASA-MIA, KASCADE and KASCADE-GRANDE, which abruptly change behaviour at high DM masses, about $10^{13}-10^{14}$~GeV. However, in this mass range, they are much weaker than the lower limits placed by PAO and TA; therefore the cumulative constraints are independent of this feature. In connection with this, we remind the reader that we obtained these limits considering only the prompt component, which we expect to be the dominant one. This is a conservative choice, since accounting also for the secondary gamma-ray production might lead to more stringent constraints, which, however, would be more dependent on the properties of the target photons for secondary production (e.g. SL+IR light). Furthermore, as we will see below, the comparison with Ishiwata et{\color{black}.} al. shows that at least for the $b\bar{b}$ channel we can recover the results obtained using also the secondary component.

\begin{figure}[t!]
    \centering
    \includegraphics[width=\textwidth]{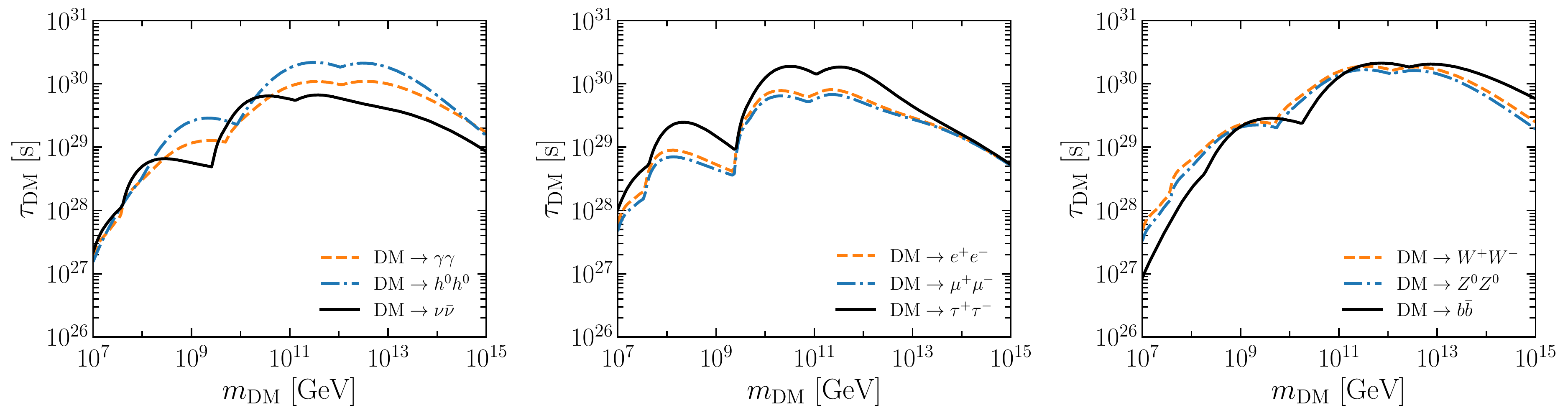}
    \caption{95\% CL lifetime limits of decaying DM particles for a wide variety of decaying channels. We show only the upper limits placed by the combination of the limits of each experimental collaboration. In each panel we show three different decay channels.}\label{fig:DM_life_allchannels}.
\end{figure}

We now extend this analysis to 6 extra decay channels ($W^+W^-$, $e^+e^-$, $Z^0Z^0$, $\gamma\gamma$, $\mu^+\mu^-$, $h^0h^0$). The total upper limits for each decay channel are shown in Fig.~\ref{fig:DM_life_allchannels}, where the combination of the KASCADE and PAO limits provide the upper constraints. In this figure we can see that the leptonic, hadronic, and bosonic limits present a rather similar behaviour respectively. 

\begin{figure}
    \centering
    \includegraphics[width=\textwidth]{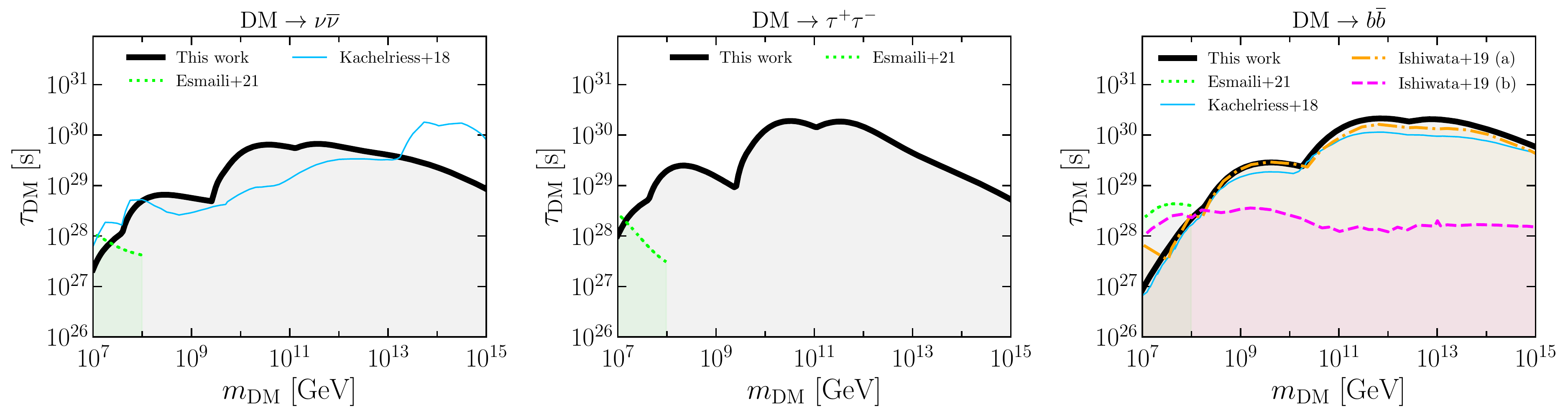}
    \caption{Comparison of the upper constraints obtained in this work (solid black line) with the complementary gamma-ray existing limits. Each column represents a DM decay channel. We show the current picture on the gamma-ray constraints of the lifetime of heavy decaying DM, where the green dotted line correspond to the limits placed by the angular analysis of the Tibet-AS data performed in Esmaili et. al.~\cite{Esmaili:2021yaw}. Also, in the $b\bar{b}$ channel we introduce the (a) galactic ({\color{black}orange} dot-dashed) and (b) extragalactic (pink dashed) limits placed by Ishiwata et. al.~\cite{Ishiwata:2019aet}, where we can see the excellent agreement between our limits and their galactic ones. Finally, we also show the gamma-ray limits placed by Kalcheriess et. al.~\cite{Kachelriess:2018rty} {\color{black}(cyan solid)} on the $\nu\bar{\nu}$ and $b\bar{b}$ channels using the same experimental measurements than the ones used in this work.}\label{fig:MM_gammapicture}
\end{figure}

We show in Fig.~\ref{fig:MM_gammapicture} the comparison between our limits and the previous ones in the literature using gamma-rays. We show three channels: $\nu\bar{\nu}$, $\tau^+\tau^-$, and $b\bar{b}$, because the $\nu\bar{\nu}$ and $b\bar{b}$ channels are the only ones with existing constraints for this DM masses. The $\tau^+\tau^-$ channel is also shown as a representative for all the other ones, for which these constraints are obtained here for the first time in this range of masses.
We also include the lifetime limits obtained from the measurements of Tibet-AS$\gamma$ by Esmaili et. al.~\cite{Esmaili:2021yaw}, shown as a dotted green line on the three panels. As we can see they cover a very small low-mass range, being complementary constraints for $10^7-10^8$ GeV DM masses.
If we focus on the $b\bar{b}$ channel we find that the dot-dashed blue line and the dashed pink one are the galactic and extragalactic multi-messenger constraints placed by Ishiwata et{\color{black}.} al.~\cite{Ishiwata:2019aet}, respectively. The excellent agreement between their galactic constraints and the ones derived in this work validates the assumption that we initially made about the prompt component being the dominant one. The extragalactic constraints surpass the galactic ones for low DM masses, but in this region the limits obtained by the angular analysis of the Tibet-AS$\gamma$ data are the dominant ones.
Finally, the results obtained by Kachelriess et. al.~\cite{Kachelriess:2018rty} place similar constraints to the galactic ones obtained by Ishiwata et. al., and therefore the ones placed in this work.
However, for the $\nu\bar{\nu}$ decay channel the Kalcheriess et. al. lifetime limits behaviour is quite different from the one obtained in this work. This stems from the difference in the approach to the gamma-ray production from DM decay, translating into different fluxes and therefore inconsistent lifetime limits. This difference is more relevant for {\color{black}the} neutrinophilic channel, for which the different treatment of electroweak corrections is essential. We mention again that in this work we rely on the gamma-ray spectra from DM decay generated by \texttt{HDMSpectra}~\cite{Bauer:2020jay}.
Finally, we remark again that on all the other channels there are no pre-existing constraints, so that in Fig.~\ref{fig:DM_life_allchannels} we place the first ones for the remaining channels.

\begin{figure}
    \centering
    \includegraphics[width=\textwidth]{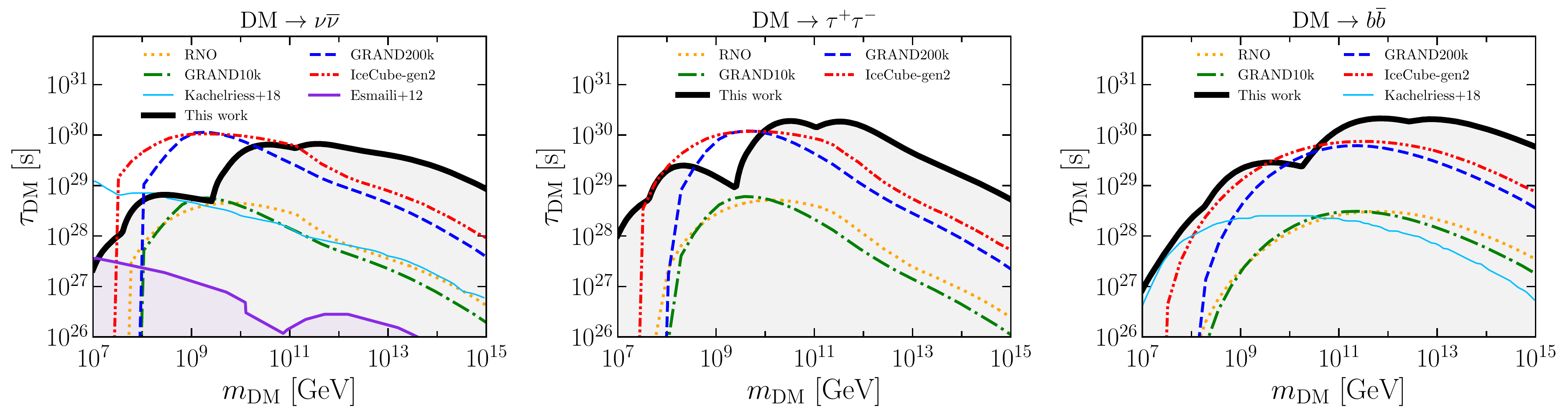}
    \caption{Upper lifetime limits comparison using neutrino detection and the limits placed by this work using gamma-rays (solid black line). We also include the projected limits that the radio neutrino telescopes may be able to place in an observation window of 3 years~\cite{Chianese:2021htv}. Each column represents a DM decay channel. In the $\nu\bar{\nu}$ channel we also have in purple solid the constraints placed by IceCube+PAO+ANITA that are presented by Esmaili et. al.~\cite{Esmaili:2012us} {\color{black}and in cyan the constraints placed by Kachelriess et. al.~\cite{Kachelriess:2018rty}}.}\label{fig:MM_nufuture}
\end{figure}

To conclude this section, we compare the result of this work with the current lifetime limits placed by neutrino observations and also the future radio telescopes projected ones obtained in our previous work~\cite{Chianese:2021htv}. This comparison is done in Fig.~\ref{fig:MM_nufuture}, where the black lines represent the limits obtained in this work. {\color{black}We also report the existing neutrino limits placed by Esmaili et. al.~\cite{Esmaili:2012us} (solid purple line) and Kachelriess et. al.~\cite{Kachelriess:2018rty} (solid cyan line). As can be seen, these limits are generally weaker than the ones placed in this work, meaning that the current upper limits are placed by gamma-ray measurements.} In order to compare with the future limits that we obtained in~\cite{Chianese:2021htv} we take into account four different neutrino radio telescopes that will try to detect EeV neutrinos: IceCube-Gen2 radio, GRAND200k, GRAND10k, and RNO-G. The main problem at these energies is that the neutrino flux is unknown and we only have theoretical predictions of the flux. In Fig.~\ref{fig:MM_nufuture} we show the \textit{optimistic} limits that will be placed by the aforementioned four different neutrino telescopes assuming that they do not observe any astrophysical neutrino in 3 years of exposure time. Focusing on the $b\bar{b}$ channel we can see that the 3-year expected limits placed by the neutrino telescopes are generally lower than the ones placed using gamma-ray measurements. In this case the neutrino limits will be complementary to the gamma-ray ones but in principle they will not overcome them. Also, the bigger telescopes are the ones that will start later taking data, so in principle we could have improvements in the gamma-ray measurements during these following years. Therefore, in principle, for the $b\bar{b}$ channel the future neutrino radio telescopes will mainly provide an independent way of obtaining complementary constraints to the gamma-ray ones. If we focus on the $\tau^+\tau^-$ and especially the $\nu\bar{\nu}$ channels, we can see that in principle for high DM masses the situation is the same as in the $b\bar{b}$ channel. However, for lower masses we find that there is a possibility to constrain a new region of the parameter space. Since the higher limits are placed by the latest constructed telescopes it is possible that the future gamma-ray measurements will exclude the parameter space that is shown on the first two panels of Fig.~\ref{fig:MM_nufuture}, but we will really need to wait and see how the gamma-ray and neutrino measurements complement each other, having neutrino telescopes the potential to place the upper constraints in the low-mass range ($10^7-10^{11}$)~GeV.

\section{Conclusions}\label{sec:conclusions}

Current gamma-ray measurements are a very powerful tool to test the lifetime of heavy dark matter particles. In this work we have used the current high-energy gamma-ray measurements in order to place constraints on the lifetime of heavy decaying dark matter particles. In particular, we have used the measurements obtained by KASCADE, KASCADE-Grande, CASA-MIA, PAO, and TA, and we have placed conservative constraints on the lifetime of heavy dark matter with masses within the range of $10^7-10^{15}~\mathrm{GeV}$. We have analyzed nine different decay channels which correspond to hadronic-philic ($b\bar{b}$), leptophilic ($\tau^+\tau^-$, $\mu^+\mu^-$, $e^+e^-$), bosonic-philic ($\gamma\gamma$, $W^+W^-$, $Z^0Z^0$, $h^0h^0$) and neutrinophilic ($\nu\bar{\nu}$) dark matter particles.

We place the upper current limits for most of the mass range $10^7-10^{15}$~GeV. Our results have been obtained taking into account only the prompt contribution, which makes them conservative, since including any other contribution to the gamma-ray flux will translate into higher constraints on the lifetime. For the $b\bar{b}$ channel we reproduced the galactic results by Ishiwata et. al., which is a good check of the dominance of the prompt contribution. For the $\nu\bar{\nu}$ channel we obtained higher constraints than the ones by Esmaili et. al. {\color{black}and Kachelriess et. al.}, and for all the other channels under study we are the first to place constraints at such high energies. Finally, the future  neutrino radio telescopes are expected to complement the gamma-ray results, providing more robustness to the constraints and possibly leading to stronger limits for some of the channels, such as $\nu\bar{\nu}$ and $\tau^+\tau^-$.

\section*{Acknowledgements}
This work was partially supported by the research grant number 2017W4HA7S ``NAT-NET: Neutrino and Astroparticle Theory Network'' under the program PRIN 2017 funded by the Italian Ministero dell'Universit\`a e della Ricerca (MUR). The authors also acknowledge the support by the research project TAsP (Theoretical Astroparticle Physics) funded by the Istituto Nazionale di Fisica Nucleare (INFN). {\color{black}RH is also supported by the Spanish grant FPU19/03348 of MU.}

\bibliography{references}

\end{document}